\pdfoutput=1
\documentclass[pra,aps,superscriptaddress,amsmath,amssymb,twocolumn,nofootinbib]{revtex4-1}

\usepackage{braket,verbatim,bm,bbold,amsmath,amssymb,epsfig,float,color}
\usepackage[colorlinks,bookmarks=false,citecolor=red,linkcolor=blue,hyperfootnotes=true,urlcolor=magenta]{hyperref}
\usepackage{color}
\usepackage[dvipsnames]{xcolor}
\usepackage{enumitem}   
\usepackage{amsmath,amsfonts,amssymb,stackengine,graphicx}
\setstackgap{S}{-1.5pt}

\newcommand{\nc}{\newcommand}
\nc{\ir}{\mathrm{i}}
\nc{\dd}{\mathrm{d}} 
\nc{\eE}{\mathsf{e}}
\nc{\Tr}{\text{Tr}}
\nc{\id}{\mathbb{I}}
\nc{\I}{\mathcal{I}}
\nc{\F}{\mathcal{F}}
\nc{\A}{\mathcal{A}}
\nc{\B}{\mathcal{B}}
\newcommand{\rrangle}{\rangle\!\rangle}
\newcommand{\llangle}{\langle\!\langle}

\newcommand{\appropto}{\mathrel{\vcenter{
  \offinterlineskip\halign{\hfil$##$\cr
    \propto\cr\noalign{\kern2pt}\sim\cr\noalign{\kern-2pt}}}}}

\begin{document}

\title{Multipartite information of free fermions on Hamming graphs}

\author{Gilles Parez}
\email{gilles.parez@umontreal.ca}
\affiliation{\it Centre de Recherches Math\'ematiques (CRM), Universit\'e de Montr\'eal, P.O. Box 6128, Centre-ville
Station, Montr\'eal (Qu\'ebec), H3C 3J7, Canada}

\author{Pierre-Antoine Bernard}
\email{pierre-antoine.bernard@umontreal.ca}
\affiliation{\it Centre de Recherches Math\'ematiques (CRM), Universit\'e de Montr\'eal, P.O. Box 6128, Centre-ville
Station, Montr\'eal (Qu\'ebec), H3C 3J7, Canada}

\author{Nicolas Cramp\'e}
\email{crampe1977@gmail.com}
\affiliation{\it  Institut Denis-Poisson CNRS/UMR 7013, Universit\'e de Tours - \\ Universit\'e d'Orl\'eans, Parc de Grandmont, 37200 Tours, France}

\author{Luc Vinet}
\email{luc.vinet@umontreal.ca}
\affiliation{\it Centre de Recherches Math\'ematiques (CRM), Universit\'e de Montr\'eal, P.O. Box 6128, Centre-ville
Station, Montr\'eal (Qu\'ebec), H3C 3J7, Canada}
\affiliation{\it IVADO, 6666 Rue Saint-Urbain, Montr\'eal (Qu\'ebec), H2S 3H1, Canada}

\date{\today}

\begin{abstract}
We investigate multipartite information and entanglement measures in the ground state of a free-fermion model defined on a Hamming graph. Using the known diagonalization of the adjacency matrix, we solve the model and construct the ground-state correlation matrix. Moreover, we find all the eigenvalues of the chopped correlation matrix when the subsystem consists of $n$ disjoint Hamming subgraphs embedded in a larger one. These results allow us to find an exact formula for the entanglement entropy of disjoint graphs, as well as for the mutual and tripartite information. We use the exact formulas for these measures to extract their asymptotic behavior in two distinct thermodynamic limits, and find excellent match with the numerical calculations. In particular, we find that the entanglement entropy admits a logarithmic violation of the area law which decreases the amount of entanglement compared to the area law scaling. 
\end{abstract}

\maketitle
\tableofcontents
\section{Introduction}
Quantum entanglement plays a prominent role in our understanding of collective phenomena in quantum many-body systems \cite{amico2008entanglement,laflorencie2016quantum}, such as quantum phase transitions \cite{vidal2003entanglement,CC04}, the emergence of thermodynamics out of equilibrium \cite{cc-05,ac-17} or topological phases of matter \cite{kitaev2006topological,levin2006detecting}.

For a quantum system in a pure state $|\psi\rrangle$, the entanglement between a subsystem $\A$ and its complement, traditionally denoted $\B$, can be quantified by the entanglement entropy. It is defined as the von Neumann entropy of the reduced density matrix $\rho_\A$ of subsystem~$\A$, 
\begin{equation}
    S(\A) = -\Tr( \rho_\A \log \rho_\A), \qquad \rho_\A = \Tr_\B(|\psi\rrangle \llangle \psi |),
\end{equation}
where $\Tr_\B$ denotes the partial trace over the degrees of freedom of $\B$. 

The entanglement entropy measures the entanglement between $\A$ and $\B$, irrespective of the geometry of $\A$. When $\A=\bigcup_{j=1}^n \A_j$ consists of $n$ disjoint subsystems, the entanglement entropy does not provide information on the multipartite entanglement between the parts of~$\A$. In the case $n=2$ where $\A$ is a bipartite subsystem, one often considers the mutual information $I_2(\A_1:\A_2)$, defined as
\begin{equation}
\label{eq:infoMut}
I_2({\A_1:\A_2}) = S({\A_1})+S({\A_2})-S({\A_1\cup \A_2}).
\end{equation}
We stress that the mutual information is not \textit{per se} an entanglement measure, because it also contains classical correlations between $\A_1$ and $\A_2$ \cite{wvhc-08}, and a proper measure of entanglement in that case is instead the entanglement negativity \cite{vw-02}. However, the mutual information shares important properties with the negativity \cite{ac2-19,bertini2022entanglement,liu2022entanglement}, and we focus here on the mutual information for simplicity. Moreover, since the mutual information measures both classical and quantum correlations, a vanishing or subleading mutual information is sufficient to indicate negligible entanglement between the subsystems.

For the cases $n>2$ where $\A$ is a multipartite subsystem, one can also define measures that quantify the presence of multipartite entanglement and correlations between the subsystems. Developing a refined understanding of multipartite entanglement and information in quantum many-body systems is a timely challenge which generates an intense theoretical activity \cite{hosur2016chaos,schnaack2019tripartite,sunderhauf2019quantum,Berthiere:2020ihq,Liu:2021ctk,tam2022topological,parez2022separability,carollo2022entangled,maric2022universality}. For $n=3$, we use the tripartite information $I_3(\A_1:\A_2:\A_3)$, defined as \cite{cerf1998information}
\begin{multline}
\label{eq:infoTri}
I_3({\A_1:\A_2:\A_3}) =\\ I_2(\A_1:\A_2)+I_2(\A_1:\A_3)-I_2(\A_1:(\A_2 \cup \A_3)).
\end{multline} 
This quantity measures the extensiveness of the mutual information. In particular, a negative tripartite information indicates multipartite entanglement which is related to quantum chaos and scrambling \cite{hosur2016chaos,schnaack2019tripartite,sunderhauf2019quantum}. In the context of two-dimensional systems where $\A_1,\A_2,\A_3$ are adjacent regions, the tripartite information coincides with the celebrated topological entanglement entropy \cite{kitaev2006topological}. Very recently, the tripartite information was investigated in the context monitored spin chains \cite{carollo2022entangled} and quantum quenches \cite{maric2022universality,pb22}. 

The mutual and tripartite information simplify in the case where the subsystems $\A_j$ are complementary, namely when $\B=\emptyset$. In that case, for $n=2$ we have $S({\A_1\cup \A_2})=0$ and $S({\A_1})=S(\A_2)$, which implies
\begin{equation}
\label{eq:I2Compl}
    I_2({\A_1:\A_2}) =2 S({\A_1}), \qquad \B=\emptyset.
\end{equation}
Similarly, for $n=3$ it is direct to show that the tripartite information vanishes,
\begin{equation}
\label{eq:I3Compl}
    I_3({\A_1:\A_2:\A_3})=0, \qquad \B=\emptyset.
\end{equation}

Any lattice quantum many body system is naturally defined on a graph: the vertices are the spatial position available for the particles, and the edges indicate the interactions. Typical examples include the one-dimensional chain and the square lattice in two dimensions. However, understanding quantum systems whose degrees of freedom reside on the vertices of more intricate graphs has received substantial attention, both from the quantum information and graph theory communities~\cite{hein2004multiparty,christandl2005perfect,guhne2005bell,hein2006entanglement,markham2008graph}. A natural endeavor in this context is to use the mathematical structure of the underlying graphs to compute entanglement measures of such quantum systems~\cite{hein2004multiparty,hein2006entanglement,calabrese2012entanglement,jafarizadeh2015investigation,crampe2020entanglement,BCV21,bernard2021entanglement,bernard2022entanglement}. 

In this paper, we investigate the multipartite entanglement properties of free fermions defined on Hamming graphs, which are natural generalizations of hypercubes in arbitrary dimension. These graphs arise in algebraic combinatorics as part of the Hamming scheme, an example of $P$- and $Q$ polynomial association schemes \cite{bannai2021algebraic,brouwer2012distance} with $su(2)$ as its Terwilliger algebra \cite{terwilliger1992subconstituent,terwilliger1993subconstituent,terwilliger1993subconstituentB,go2002terwilliger}. Physically, Hamming graphs are known to admit perfect state transfer \cite{christandl2005perfect} and fractional revival \cite{bernard2018graph}. The entanglement entropy of free fermions defined on Hamming graphs has been investigated in Ref. \cite{BCV21}, and we expand these results to the case of multipartite subsystems. 

The paper is organized as follows. In Sec. \ref{sec:II} we recall the definition of Hamming graphs and the diagonalization of their adjacency matrix. We use these results to define and diagonalize the related free-fermion Hamiltonian and construct the chopped correlation matrix. We give the exact results for the entanglement entropy of~$n$ disjoint Hamming subgraphs embedded in a larger one in Sec. \ref{sec:III}, and also provide analytical and numerical results for the asymptotic behavior of the entropy in the limit of large systems. In Sec. \ref{sec:IV} we derive exact formulas and asymptotics for the mutual and tripartite information. We offer a summary of our results and an outlook for future work in Sec. \ref{sec:V}. The diagonalization of the chopped correlation matrix is carried out in App. \ref{app:CA}.

\section{Model and definitions} \label{sec:II}
\subsection{Hamming graph} \label{sec:hamm}
The Hamming graph $H(d,q)$ is defined as follows. The set of vertices $F_q^d$ consists of all the $d$-tuples $v=(v_1,v_2,\dots,v_d)$ with $v_i \in \{0,1,\dots,q-1\}$ for all $i=1,2,\dots,d$. Two vertices $v,v' \in F_q^d$ are connected by an edge if there is exactly one position $i$ for which $v_i \neq v'_i$. The graph distance between two vertices $v,v'$, denoted $\partial (v,v')$, is defined as the number of entries for which they differ,
\begin{equation}
    \partial (v,v') = \big|\{ i \in \{1,2,\dots,d \} : v_i \neq v'_i \} \big|,
\end{equation}
where $| \{ \dots \}|$ denotes the cardinality of the set. By definition, the maximal distance between two vertices is $d$, and for this reason we say that $d$ is the diameter of the graph. 

The adjacency matrix $A$ of the Hamming graph is a $q^d \times q^d$ matrix whose entries are labeled by the vertices of~$H(d,q)$, and the components are
\begin{equation}
    [A]_{v,v'} = \left\{
\begin{array}{cc}
1 & \text{if} \ \partial (v,v')=1, \\
0 & \text{otherwise}.
\end{array}
\right.
\end{equation}

The diagonalization of $A$ is given in Ref. \cite{BCV21,Schwenk1974}, and we recall the main steps. Each vertex $v=(v_1,v_2,\dots,v_d)\in F_q^d$ can be represented by a vector in~$(\mathbb{C}^q)^{\otimes d}$,
\begin{equation}
    |v\rangle = |v_1\rangle \otimes |v_2\rangle \otimes \cdots \otimes |v_d\rangle,
\end{equation}
where 
\begin{equation}
\label{eq:vi}
    |v_i \rangle = (\underbrace{0,0,\dots,0}_{v_i \ \text{times}},1,0,\dots,0)^T.
\end{equation}

In this basis, the adjacency matrix reads 
\begin{equation}
\label{eq:A_IJ}
    A  = \sum_{i=1}^d (\mathbb{1}_{q \times q})^{\otimes i-1}\otimes (J_{q \times q}-\mathbb{1}_{q \times q}) \otimes  (\mathbb{1}_{q \times q})^{\otimes d-i}
\end{equation}
where $\mathbb{1}_{q \times q}$ is the $q \times q$ identity matrix, and $J_{q \times q}$ is the $q \times q$ matrix filled with ones. 

The matrix $J_{q \times q}$ has two distinct eigenvalues, $q$ and $0$, with respective degeneracies $1$ and $(q-1)$. The unique normalized eigenvector associated to the eigenvalue $q$ is 
\begin{equation}
    | \theta_q \rangle  = \frac{1}{\sqrt{q}} \sum_{i=0}^{q-1} |i\rangle,
\end{equation}
whereas the $q-1$ orthonormal eigenvectors with zero eigenvalue are denoted $|\theta_j \rangle$, $j=1,2,\dots,q-1$. The set $\{|\theta_i\rangle  : i \in \{1,2,\dots q\}\}$ thus defines an orthonormal basis of~$\mathbb{C}^q$. 

It follows that eigenvectors of $A$ are of the form $|\theta_{i_1} \theta_{i_2} \dots \theta_{i_d}\rangle \equiv |\theta_{i_1}\rangle \otimes |\theta_{i_2}\rangle \otimes \cdots \otimes |\theta_{i_d}\rangle$ with $i_j \in \{1,2,\dots,q\}$. Indeed, from Eq. \eqref{eq:A_IJ} we have 
\begin{equation}
    A |\theta_{i_1} \theta_{i_2} \dots \theta_{i_d}\rangle = \omega_k|\theta_{i_1} \theta_{i_2} \dots \theta_{i_d}\rangle,
\end{equation}
where $k$ is the number of vectors $|\theta_q\rangle$ in the tensor product. The eigenvalue is 
\begin{equation}
\label{eq:omega}
    \omega_k \equiv kq-d, \qquad k=0,1,\dots,d,
\end{equation}
and it has the degeneracy 
\begin{equation}\label{eq:Dk}
    D_k = \begin{pmatrix}
    d \\ k
\end{pmatrix}(q-1)^{d-k}.
\end{equation}
In the following, we denote the orthonormal eigenvectors of $A$ as $|\omega_k,m_k\rangle$, where $\omega_k$ is the eigenvalue and $m_k=1,2,\dots,D_k$ labels the degeneracies.

\subsection{Free-fermion Hamiltonian and ground state}

We consider spinless free fermions hopping between vertices of the Hamming graph $H(d,q)$ with distance-dependent amplitudes $\alpha_i \in \mathbb{R}$, $i=0,1,\dots,d$. Physically, $\alpha_0$ is the chemical potential, $\alpha_1$ is the nearest-neighbor hoping amplitude, and $\alpha_{i>1}$ correspond to long-range interactions. The Hamiltonian is
\begin{equation}
    \mathcal{H}= \sum_{v,v' \in F_q^d} \alpha_{\partial (v,v')} c^{\dagger}_v c_{v'}
\end{equation}
where $c^\dagger_v,c_v$ are fermionic creation and annihilation operators satisfying the canonical anticommutation relation
\begin{equation}
    \{c^\dagger_v,c_{v'}\} = \delta_{v,v'}, \quad  \{c^\dagger_v,c^\dagger_{v'}\}=  \{c_v,c_{v'}\}=0.
\end{equation}

The Hamiltonian acts on a Hilbert space of dimension $2^{|F_q^d|}$, where $|F_q^d| =q^d$ is the number of vertices in the Hamming graph $H(d,q)$. The vacuum state, denoted $|0\rrangle$, is a special vector of this Hilbert space which is annihilated by all fermionic annihilation operators, $c_v |0\rrangle=0 \ \forall v \in F_q^d$.  

We diagonalize $\mathcal{H}$ using the diagonalization of the adjacency matrix $A$ presented in the previous section. 
We introduce the diagonal fermionic operators $d^\dagger_{k,m_k}, d_{k,m_k}$,
\begin{equation}
    d^\dagger_{k,m_k} = \sum_{v \in F_q^d} \langle v|\omega_k,m_k \rangle  c^\dagger_v, \quad d_{k,m_k} = \sum_{v \in F_q^d} \langle \omega_k,m_k | v \rangle  c_v,
\end{equation}
and recast the Hamiltonian in diagonal form,
\begin{equation}
    \mathcal{H} = \sum_{k=0}^d \sum_{m_k=1}^{D_k} \epsilon_k \ d^\dagger_{k,m_k} d_{k,m_k}.
\end{equation}

The single-particle energies $\epsilon_k$ are \cite{BCV21}
\begin{equation}
\label{eq:epsilon}
    \epsilon_k = \sum_{i=0}^d \alpha_i\begin{pmatrix}
    d \\ i
\end{pmatrix}(q-1)^{i}K_i \Big(d-k;\frac{q-1}{q},d\Big)
\end{equation}
where $K_i$ are the Krawtchouk polynomials of degree~$i$~\cite{koekoek2010hypergeometric},
\begin{equation}
K_i \Big(d-k;\frac{q-1}{q},d\Big) = \sum_{j=0}^i \frac{(-i)_j(k-d)_j}{(-d)_j \ j!}\left(\frac{q}{q-1}\right)^j.
\end{equation}
Here, $(a)_j$ is the Pochammer symbol (or shifted factorial), defined as
\begin{equation}
    (a)_0=1, \qquad (a)_j = a(a+1)\cdots(a+j-1), \quad j>0.
\end{equation}

From this diagonalization, one can construct all the eigenstates of $\mathcal{H}$ by applying diagonal creation operators on the vacuum state $|0\rrangle$. In particular, the  ground state~$|\psi_0\rrangle$ is
\begin{equation}
\label{eq:psi0}
    |\psi_0 \rrangle  = \prod_{k \in \F} \prod_{m_k=1}^{D_k} d^\dagger_{k,m_k} |0\rrangle
\end{equation}
where $\F$ is the set of all integers $k \in \{0,1,\dots,d\}$ such that $\epsilon_k \leqslant 0$. We note that the ground state is degenerate if there are vanishing single-particle energies $\epsilon_{k_0}=0$.

\subsection{Nearest-neighbor and long-range models}

In this section, we give two example of free-fermion models defined on Hamming graphs for which there is a closed-form formula for the energies \eqref{eq:epsilon}. Moreover, the set $\F$ has the simple form $\F=\{0,1,\dots, k_0 \}$ where $k_0$ depends on the specific model.

As a first example, let us consider the model nearest-neighbor hoping, namely $\alpha_1=1$ and $\alpha_{i>1}=0$. In this case, Eq. \eqref{eq:epsilon} simplifies greatly, and a direct calculation yields 
\begin{equation}
\label{eq:epsilon01}
    \epsilon_k = \alpha_0 + \omega_k ,
\end{equation}
where $\omega_k$ are the eigenvalues of the adjacency matrix, see Eq. \eqref{eq:omega}. This is the generalization to the Hamming graph of the well-known diagonalization of the one-dimensional tight-biding model \cite{lieb1961two}. In this situation, the Fermi momentum is 
\begin{equation}
\label{eq:koNN}
k_0 = \left \lfloor\frac{d-\alpha_0}{q} \right \rfloor .
\end{equation}
This model has a finite energy gap $q$, and the degeneracy of the first excited state is exponentially large in $d$.

It is also possible to find a closed-form expression for the energies in the case of long-range interactions with exponential suppression, namely $\alpha_{i>0}=\eE^{-c i}$ with $c\geqslant 0$. The energies in this case read \cite{BCV21} 
\begin{equation}
\label{eq:ekLR}
\epsilon_k = (1-\eE^{-c})^{d-k}(1+\eE^{-c}(q-1))^{k}+\alpha_0-1.
\end{equation}
For $\alpha_0\geqslant 1$ there are no negative energies and $\F=\emptyset$, whereas for $\alpha_0<1$ we have
\begin{equation}
\label{eq:koLR}
k_0  = \left \lfloor\frac{\log(1-\alpha_0)-d\log(1-\eE^{-c})}{\log(1+\eE^{-c}(q-1))-\log(1-\eE^{-c})} \right \rfloor.
\end{equation}

\subsection{Filling fraction}\label{sec:nuf}

Let us investigate the filling fraction $\nu_\F$ of the ground state $|\psi_0 \rrangle$ in Eq. \eqref{eq:psi0} for the nearest-neighbor and long-range models. It is defined as the ratio between the occupation number of the ground state and the total number of vertices $|F_q^d| =q^d$, 
\begin{equation}
\nu_\F = \frac{1}{q^d}\sum_{k=0}^{k_0}\begin{pmatrix}
    d \\ k
\end{pmatrix}(q-1)^{d-k}.
\end{equation}
  
\begin{figure}
\includegraphics[width=0.45\textwidth]{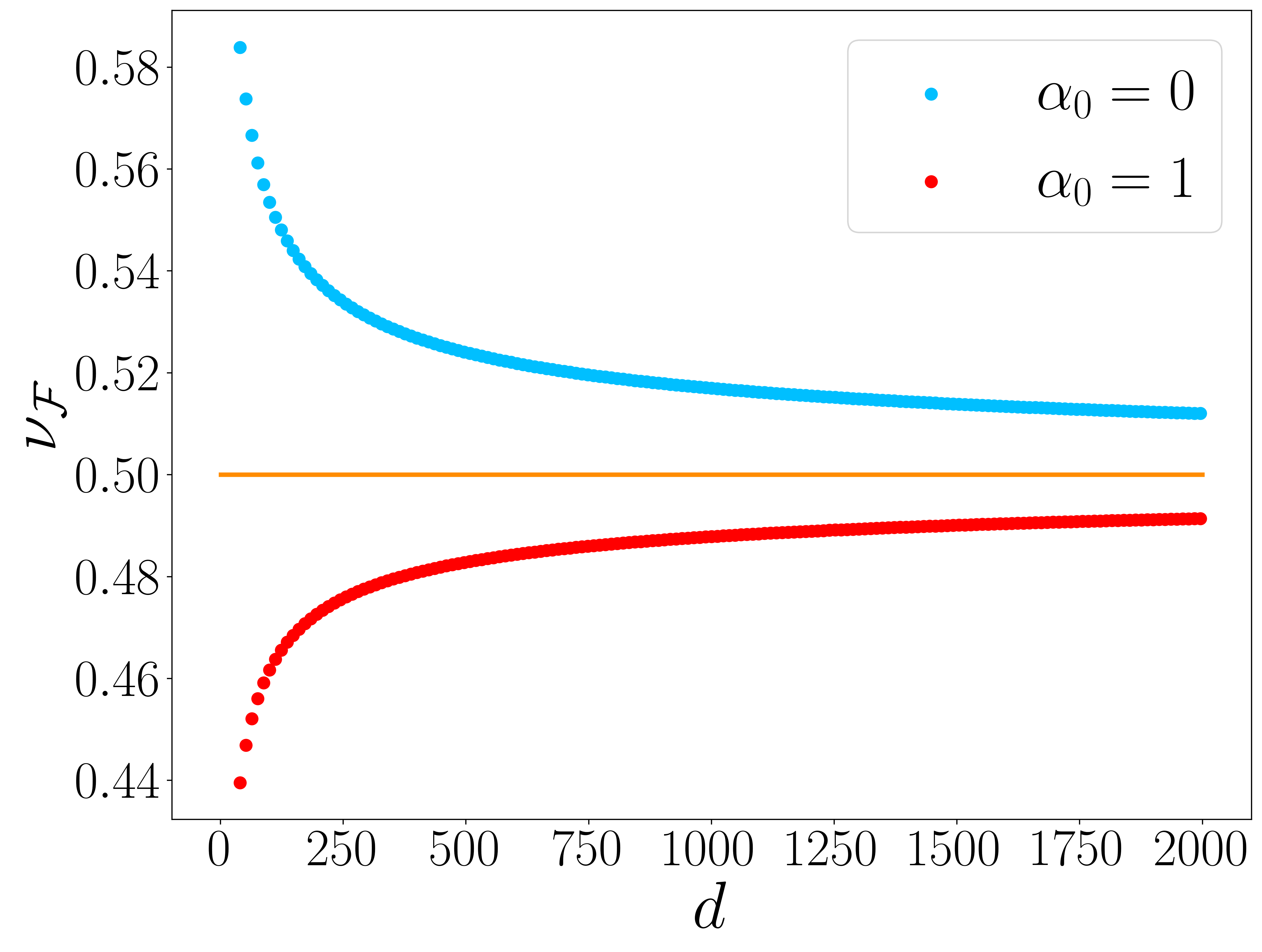}   
\caption{Ground-state filling fraction $\nu_\F$ for the nearest-neighbor hoping model with $q=4$ and $\alpha_0=0,1$ as a function of $d$ (symbols). The solid line indicates the asymptotic value $\nu_\F=1/2$.} 
\label{Fig:nuF}
\end{figure}

For the nearest-neighbor case where $k_0=\lfloor (q-\alpha_0)/q\rfloor$, we investigate the filling fraction as a function of $d$ for different value of $q$ and $\alpha_0$. Even though $\nu_\F$ is not a constant, it converges to $1/2$ for large values of $d$, irrespective of $q$ and $\alpha_0$. We illustrate this in Fig. \ref{Fig:nuF} for $q=4$ and $\alpha_0=0,1$. We have similar curves for different values of $q$ and $\alpha_0$ but do not reproduce them on the figure for clarity.  

Let us now consider the long-range model with $k_0$ defined in Eq. \eqref{eq:koLR} and investigate the filling fraction as a function of $\alpha_0$ and $c$. First, for $\alpha_0 \geqslant 1$ all the energies are positive, and therefore $\nu_\F=0$. Second, for $\alpha_0<1$ and small $c \geqslant 0$, the energies in Eq. \eqref{eq:ekLR} are all negative for large values of $d$, and therefore the filling fraction satisfies $\lim_{d\to \infty}\nu_\F=1$. Third, we consider the case $\alpha_0\neq 0$ and large $c$. In this case, the large-$d$ limit of the filling fraction depends on the sign of $\alpha_0$. We have $\lim_{d\to \infty}\nu_\F=0$ for $\alpha_0>0$ and $\lim_{d\to \infty}\nu_\F=1$ for $\alpha_0<0$. Finally, let us consider $\alpha_0=0$ and large $c$. From Eq. \eqref{eq:koLR}, we have 
\begin{equation}
k_0 \sim \left \lfloor\frac{d}{q}\right \rfloor,
\end{equation}
which is the same value as for the nearest-neighbor hoping model with $\alpha_0=0$, see Eq. \eqref{eq:koNN}. Therefore, the filling fraction converges to $1/2$, similarly as in Fig. \ref{Fig:nuF}. We conclude that for the long-range model, the filling fraction converges to either $\nu_\F=0,1$ or $1/2$ depending on $\alpha_0$ and $c$.

\subsection{Correlation matrix and entanglement entropy}

The diagonalization of $\mathcal{H}$ also allows to compute the ground-state two-point correlation matrix $C$ (or simply correlation matrix). Similarly to the adjacency matrix $A$, it is a matrix whose entries are labeled by the vertices of $H(d,q)$ and which is defined as 
\begin{equation}
    [C]_{v,v'} = \llangle \psi_0 | c^\dagger_v c_{v'} | \psi_0 \rrangle. 
\end{equation}
A direct calculation shows that 
\begin{equation}
\begin{split}
    C &= \sum_{k \in \F} \sum_{m_k=1}^{D_k}  | \omega_k,m_k \rangle \langle \omega_k,m_k | \\
   & \equiv \pi_\F,
   \end{split}
\end{equation}
namely that $C=\pi_\F$, the projection operator onto all the eigenspaces of $A$ associated to negative single-particle energies.

Let us consider a subset $\A \in F_q^d$ of vertices of the Hamming graph, and denote by $\pi_\A$ the projector on this subset. The chopped correlation matrix $C_\A$ is the restriction of the correlation matrix $C$ to the subset $\A$, i.e.
\begin{equation}
    C_\A = \pi_\A\pi_\F\pi_\A.
\end{equation}

Owing to the quadratic nature of the Hamiltonian $\mathcal{H}$, the ground-state entanglement entropy $ S(\A)$ of $\A \in F_q^d$ can be obtained via the chopped correlation matrix~\cite{peschel2003calculation,peschel2009reduced}, 
\begin{equation}
\label{eq:SAC}
    S(\A) = -\Tr (C_\A \log C_\A + (1-C_\A)\log (1-C_\A) ).
\end{equation}

\section{Entanglement entropy of disjoint subgraphs}\label{sec:III}
We consider the case where the subsystem $\A$ consists of $n$ disjoints Hamming graphs $H(L,q)$ with $L<d$, embedded in the larger one $H(d,q)$. More specifically, $\A=\bigcup_{j=1}^n \A_j$ with 
\begin{equation}
\A_j = \left \{ v \in F_q^d : v_i =(j-1)  \ \forall i \ \text{s.t.} \ 1 \leqslant i \leqslant r \right\}
\end{equation}
where $r \equiv d-L$ is an integer that we interpret as the distance between the subsystems. Indeed, the minimal path between two vertices $v \in \A_j$ and $v' \in \A_{j'}$ with $j \neq j'$ has length $r$. Moreover, we impose $n\leqslant q$. 

We note that every vertex $v \in \A$ is adjacent to $r(q-1)$ vertices in the complement $\B$. The size (or volume) of $\A$ is 
\begin{equation}
\label{eq:Va}
    V_\A = |\A| = n q^{d-r},
\end{equation}
whereas the size of the boundary (or area), defined as the number of edges connecting vertices in $\A$ and in $\B$, is 
\begin{equation}
\label{eq:VADA}
    |\partial_\A| = (q-1)r V_\A = r(q-1) \ n q^{d-r}.
\end{equation}
The fact that every vertex of $\A$ belongs to the boundary between $\A$ and $\B$ is reminiscent of the so-called skeletal regions, introduced in Ref. \cite{berthiere2022entanglement}. However, the main difference here is that the volume of $\A$ is not negligible compared to the size of the whole system. Indeed, 
\begin{equation}
\label{eq:aspectRatio}
    \frac{V_\A}{|F_q^d|} = n q^{-r},
\end{equation}
and this number is finite and non-zero for fixed $n,q,r$, even in the large-volume limit $d \to \infty$.

The entropy for one subsystem $S(\A_j)$ is computed in Ref. \cite{BCV21}. In the following, we generalize these computations to the multipartite case, and derive exact formulas for the entanglement entropies $S(\bigcup_{j=1}^n \A_j)$, which are the building blocks of the multipartite entanglement measures.  

\subsection{Eigenvectors and eigenvalues of $C_\A$}

Our goal is to construct eigenvectors of the operator $C_\A=\pi_\A \pi_\F \pi_\A$. We define 
\begin{equation}
    \eta_j = \eE^{2\ir \pi j/n},\qquad j=1,2,\dots,n.
\end{equation}
The subset $\A$ has dimension $n \times q^L$. We introduce $n \times q^L$ orthonormal vectors as follows,
\begin{equation}
|\theta_{i_1} \cdots \theta_{i_L}\rangle_{j} = \frac{1}{\sqrt{n}}\left(\sum_{\ell=0}^{n-1} \eta_j^\ell \ |\ell\rangle^{\otimes r}\right)\otimes |\theta_{i_1} \rangle \otimes \cdots \otimes |\theta_{i_L}\rangle, 
\end{equation}
where $j=1,2,\dots,n,$ and the vectors $|\theta_i\rangle$ are defined in Sec.~\ref{sec:hamm}.

Let us define $Q$ as the number of vectors $|\theta_{q}\rangle$ in the state $|\theta_{i_1} \cdots \theta_{i_L}\rangle_{j}$. We find that $|\theta_{i_1} \cdots \theta_{i_L}\rangle_{j}$ is an eigenvector of the chopped correlation matrix,
\begin{subequations}
\label{eq:CLambda}
\begin{equation}
    C_\A |\theta_{i_1} \cdots \theta_{i_L}\rangle_{j} = \left\{
\begin{array}{cc}
\Lambda_{Q,0}^{(n)}|\theta_{i_1} \cdots \theta_{i_L}\rangle_{0}, &  \ j=0, \\[.3cm]
\Lambda_{Q,1}^{(n)}|\theta_{i_1} \cdots \theta_{i_L}\rangle_{j} &  \ j>0,
\end{array}
\right.
\end{equation}
with eigenvalue
\begin{multline}
\label{eq:Lambda}
    \Lambda_{Q,e}^{(n)} = \sum_{\substack{k \in \F \\ Q \leqslant k \leqslant Q+r}} \begin{pmatrix}
        r \\ k-Q
    \end{pmatrix} \left(\frac{1}{q}\right)^{k-Q} \left(\frac{q-1}{q}\right)^{r-k+Q} \\ + \frac{n \delta_{e,0}-1}{q^r}\sum_{\substack{k \in \F \\ Q \leqslant k \leqslant Q+r}} \begin{pmatrix}
        r \\ k-Q
    \end{pmatrix} (-1)^{r-k+Q}.
\end{multline}
\end{subequations}
We give the proof of this result in App. \ref{app:CA}. For $n=1$, the second term on the right-hand side of Eq. \eqref{eq:Lambda} vanishes, and we recover the result from Ref. \cite{BCV21} for the case of a single subsystem.

 The eigenvalues $\Lambda_{Q,0}^{(n)}$ and $\Lambda_{Q,1}^{(n)}$ have degeneracy $\mathcal{D}_Q$ and $(n-1)\mathcal{D}_Q$, respectively, with $\mathcal{D}_Q$ given by Eq. \eqref{eq:Dk} with $k=Q$ and $d \rightarrow L$. The total number of eigenvalues is  
\begin{equation}
    n \sum_{Q=0}^L \mathcal{D}_Q = n q^L,
\end{equation}
as expected.

\subsection{Entanglement entropy}

We introduce the function $s(x)$ as 
\begin{equation}
    s(x) = - x\log x -(1-x)\log (1-x).
\end{equation}
With Eqs. \eqref{eq:SAC} and \eqref{eq:CLambda}, we find that the entanglement entropy of $n$ disjoint Hamming subgraphs is
\begin{equation}
\label{eq:SLambda}
    S\Big(\bigcup_{j=1}^n \A_j\Big) = \sum_{Q=0}^L \mathcal{D}_Q \Big[s(\Lambda_{Q,0}^{(n)})+(n-1)s(\Lambda_{Q,1}^{(n)})\Big].
\end{equation}

Let us consider the case where $\F=\{0,1,\dots,k_0\}$ for some Fermi momentum $k_0>0$. In that situation, if $Q>k_0$, we have $\Lambda_{Q,j}^{(n)}=0$. Moreover, for $Q \leqslant k_0-r$, the sums in Eq. \eqref{eq:Lambda} can be simplified using Newton's binomial formula, and we find $\Lambda_{Q,j}^{(n)}=1$. In both cases, the eigenvalues do not contribute to the entanglement entropy because $\lim_{x\to 0,1}s(x)=0$. Therefore, Eq. \eqref{eq:SLambda} simplifies to 
\begin{subequations}
\label{eq:SAn_exact}
\begin{multline}
    S\Big(\bigcup_{j=1}^n \A_j\Big) = \sum_{i=1}^{r} \begin{pmatrix}
    L \\ d-i-k_0
\end{pmatrix}(q-1)^{d-i-k_0}\\ \times \Big(s(F_{i,0}^{(n)})+(n-1)s(F_{i,1}^{(n)})\Big)
\end{multline}
where we introduced $i=Q-k_0+r$ and 
\begin{multline}
    F_{i,j}^{(n)} =   \sum_{m=0} ^{r-i}
\begin{pmatrix}
r \\ m
\end{pmatrix} \frac{1}{q^r} \Big( (q-1)^{r-m}+(n \delta_{j,0}-1) (-1)^{r-m}  \Big).
\end{multline}
\end{subequations}

Let us stress that the entanglement entropy on Hamming graphs only depends on the underlying model (i.e. the choice of $\alpha_i$) through $k_0$, or equivalently the filling fraction $\nu_\F$. Moreover, the entanglement entropy vanishes for the trivial filling fractions $\nu_\F=0,1$. Therefore, based on our analysis of Sec. \ref{sec:nuf}, we focus on the case $k_0=\lfloor d/q \rfloor$ which corresponds to an asymptotic filling fraction $\lim_{d\to \infty} \nu_\F=1/2$. Our results thus hold both for the nearest-neighbor hoping model with arbitrary $\alpha_0$ and the long-range model with large $c$ and $\alpha_0=0$. In the following we further assume that~$d$ is a multiple of $q$, such that $k_0=d/q$.

\subsection{Asymptotics for $r=1$}\label{sec:r1}

For $r=1$, or $L=d-1$, the sum in Eq. \eqref{eq:SAn_exact} reduces to a single term. Furthermore, for $i=r=1$, the function $F_{i,j}^{(n)} $ simplifies to 
\begin{equation}
\label{eq:Fsimple}
    F_{1,j}^{(n)} = \frac{q-n \delta_{j,0}}{q}.
\end{equation}

\begin{figure}
\includegraphics[width=0.45\textwidth]{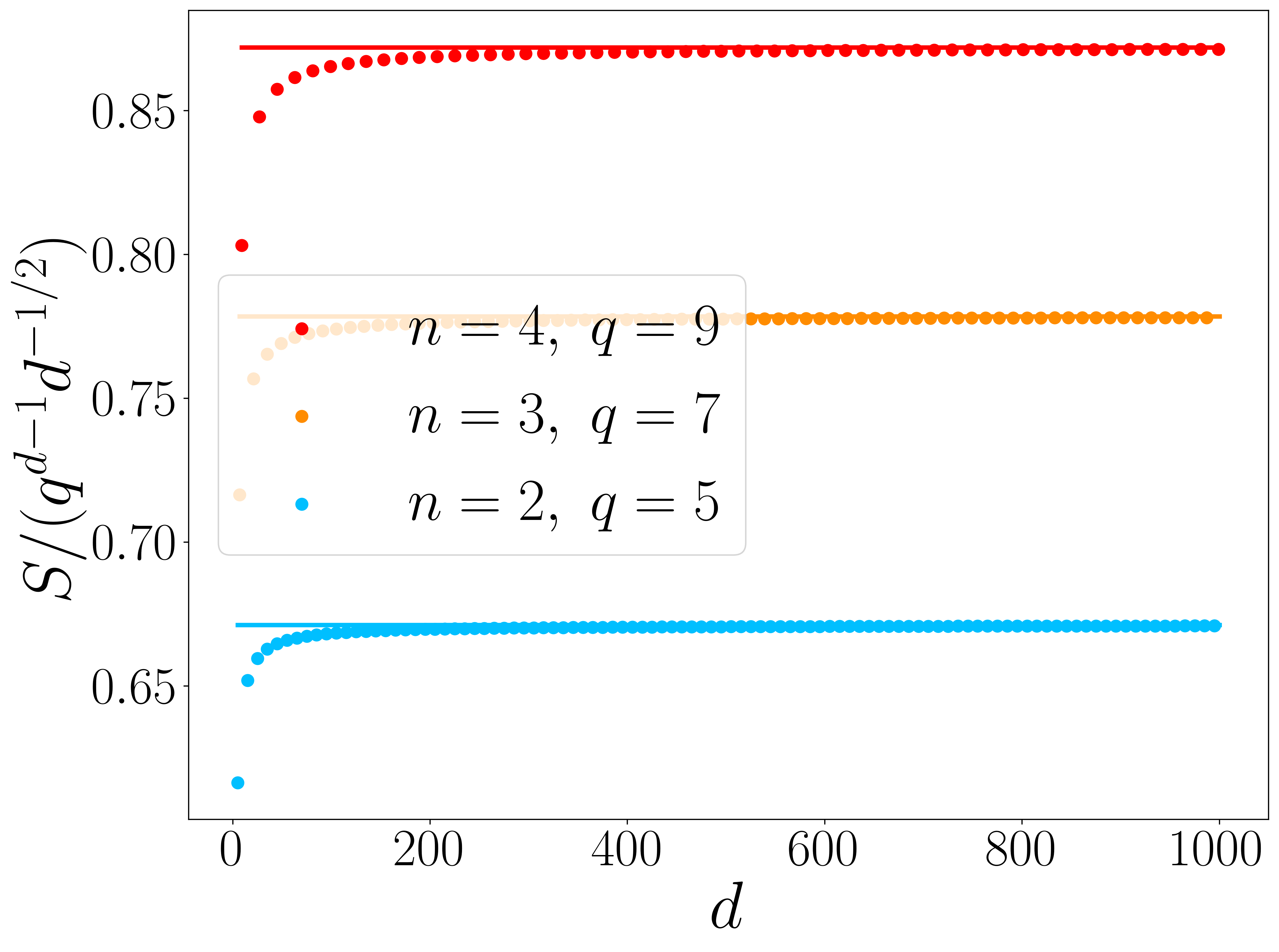}   
\caption{Entanglement entropy $S$ divided by $q^{d-1}d^{-1/2}$ as a function of $d$ for $r=1$, $k_0=d/q$, and various values of $n,q$. We compare the exact result of Eq. \eqref{eq:SAn_exact} (symbols) with the asymptotic formula of Eq. \eqref{eq:Sr1Asympt} (solid lines), and find excellent agreement.} 
\label{Fig:Sr1_Hamming}
\end{figure}

We thus have 
\begin{equation}
\label{eq:Sr1}
     S\Big(\bigcup_{j=1}^n \A_j\Big) = \begin{pmatrix}
    d-1 \\ d-1-k_0
\end{pmatrix}(q-1)^{d-1-k_0} s\Big(\frac{q-n}{q}\Big)
\end{equation}
where we used $s(0)=s(1)=0$.
%
Using Stirling's formula in Eq. \eqref{eq:Sr1}, we find
\begin{equation}
\label{eq:Sr1Asympt}
     S\Big(\bigcup_{j=1}^n \A_j\Big) \sim q^{d-1} \left(\frac{1}{d} \right)^{1/2} \frac{q}{\sqrt{2\pi (q-1)}}\ s\Big(\frac{q-n}{q}\Big)
\end{equation}
at leading order in the limit $d\to \infty$. In Fig. \ref{Fig:Sr1_Hamming}, we compare this asymptotic result (solid lines) with the exact formula of Eq. \eqref{eq:SAn_exact} (symbols), and find excellent agreement.

\subsection{Asymptotics for finite $r$}\label{sec:asymptFinr}

\begin{figure*}
\includegraphics[width=0.45\textwidth]{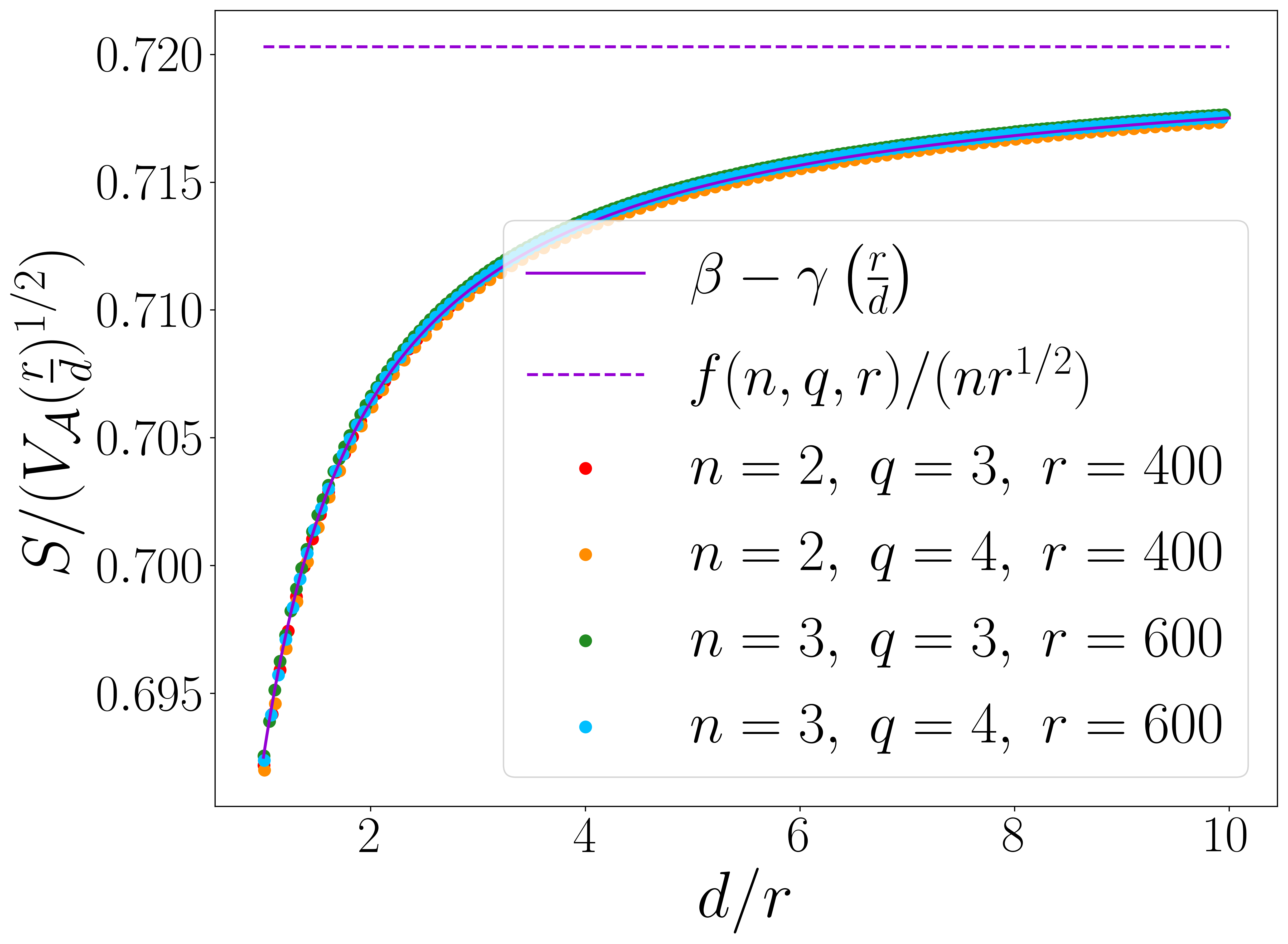}   \qquad
\includegraphics[width=0.45\textwidth]{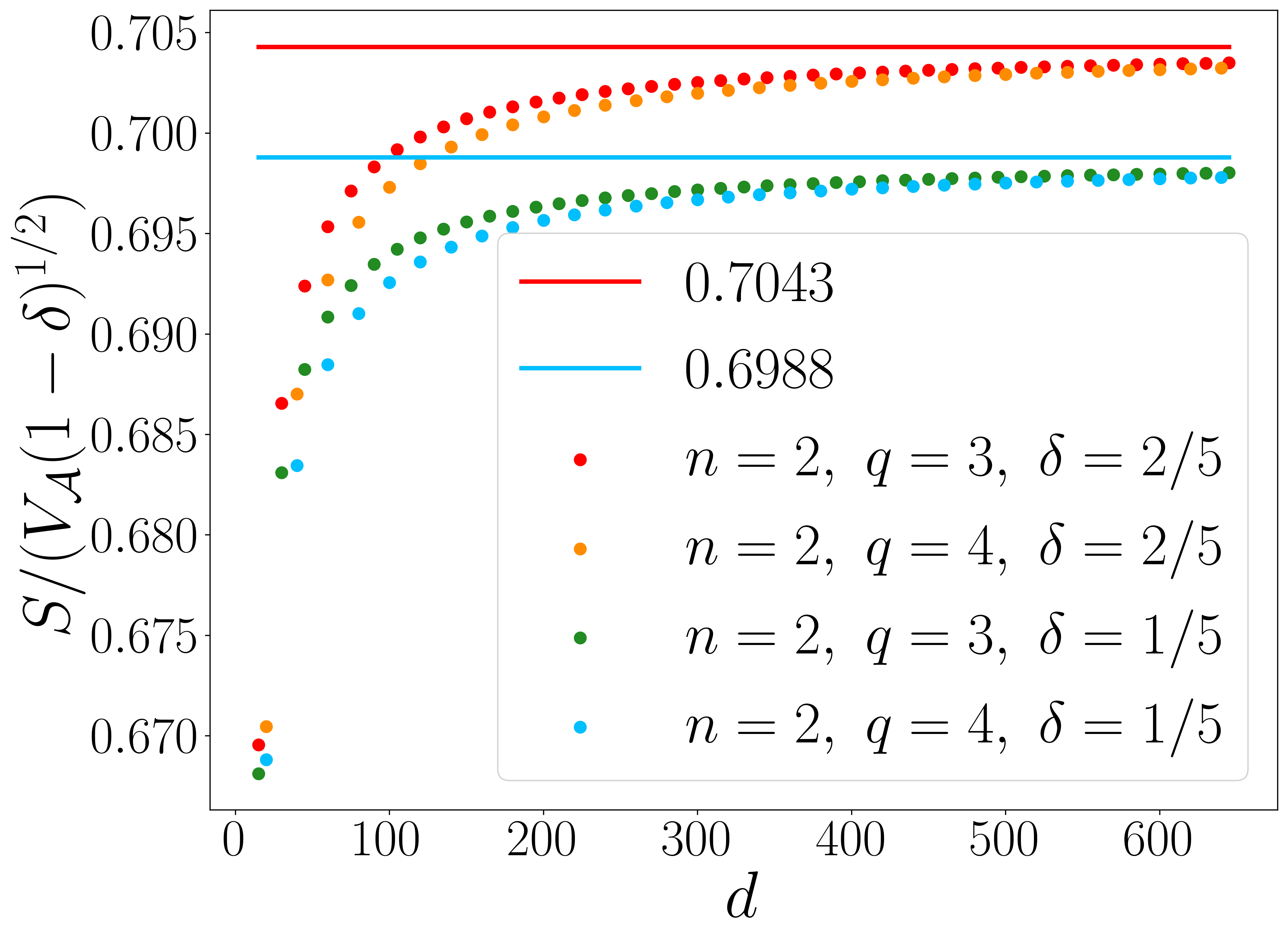}  
\caption{\textit{Left:} Entanglement entropy $S$ divided by $V_\A \left(r/d \right)^{1/2}$ as a function of $d/r$ for $k_0=d/q$ and various values of $n,q,r$. The symbols are obtained from the exact formula of Eq. \eqref{eq:SAn_exact}, and the solid line is the function $\beta- \gamma \left(\frac{r}{d} \right)$ where $\beta,\gamma$ are given in Eq.~\eqref{eq:betagamma}. The dashed line is the asymptotic value $f(n,q,r)/n r^{1/2} \sim 0.7203$ for $n=3$, $q=4$ and $r=600$, where $f(n,q,r)$ is defined in Eq. \eqref{eq:SrAsymptb}. \textit{Right:} Entanglement entropy $S$ divided by $V_\A \left(1-\delta \right)^{1/2}$ as a function of $d$ for $k_0=d/q$, $n=2$ and various values of $q,\delta$. The symbols are obtained from the exact formula of Eq. \eqref{eq:SAn_exact}, and the solid lines are the fitted asymptotic values given in Eq.~\eqref{eq:fittedTildeBeta}. } 
\label{Fig:Sn_Hamming}
\end{figure*}

It is possible to generalize the calculations of the previous section to investigate the asymptotic behavior of the entanglement entropy in the limit $d\to \infty$ where $r>0$ is fixed. This corresponds to the thermodynamic limit where the ratio ${V_\A}/{|F_q^d|}$ remains a constant, see Eq.~\eqref{eq:aspectRatio}. We find 
\begin{subequations}
\label{eq:SrAsympt}
    \begin{equation}
    \label{eq:SrAsympta}
        S\Big(\bigcup_{j=1}^n \A_j\Big) \sim q^{d-r} \left(\frac{1}{d} \right)^{1/2} f(n,q,r)
    \end{equation}
with 
\begin{equation}
\label{eq:SrAsymptb}
\begin{split}
    f(n,q,r) &= \frac{q}{\sqrt{2\pi (q-1)}} \sum_{i=1}^{r} \Big(s(F_{i,0}^{(n)})+(n-1)s(F_{i,1}^{(n)})\Big) 
    \end{split}
\end{equation}
\end{subequations}
at leading order for large $d$. As expected, Eq. \eqref{eq:SrAsympt} reduces to Eq. \eqref{eq:Sr1Asympt} for $r=1$. The agreement between the asymptotic result of Eq. \eqref{eq:SrAsympt} and the exact formula of Eq. \eqref{eq:SAn_exact} is excellent. We do not illustrate this agreement with a new figure, because it would essentially be identical to Fig. \ref{Fig:Sr1_Hamming}.  

 Even though the function $f(n,q,r)$ is cumbersome, Eq.~\eqref{eq:SrAsympta} provides the full $d$-dependence of the leading term of the entanglement entropy. Moreover, for large $r$ we numerically observe the following properties: (i) $f(n,q,r) \appropto n$, (ii)  $f(n,q,r)^2 \appropto r$, and (iii) $\partial_q f(n,q,r)\sim 0$.

These observations, combined with further numerical investigations of the entanglement entropy, suggest that the entropy scales as
\begin{subequations}
\label{eq:Snfit}
\begin{equation}
\label{eq:Snr}
    S\Big(\bigcup_{j=1}^n \A_j\Big) \sim  V_\A \left(\frac{r}{d} \right)^{1/2}\left(\beta- \gamma \left(\frac{r}{d} \right) +\dots\right) 
\end{equation}
where $V_\A= n q^{d-r}$ is given in Eq. \eqref{eq:Va}, the ellipsis indicate subleading terms of order $\left(r/d \right)^2$, and $\beta,\gamma$ are constants that do not depend on $n,q,r,d$. However, we stress that the entanglement entropy of $n$ disjoint Hamming subgraphs is not exactly proportional to $n$. Therefore, there are also subleading corrections in Eq. \eqref{eq:Snr} which are not proportional to $n$. We fit the values of $\beta,\gamma$ and find
\begin{equation}
\label{eq:betagamma}
    \beta \simeq 0.7203, \qquad \gamma \simeq 0.0278,
\end{equation}
\end{subequations}
in the limit $d \to \infty$ for large but fixed values of $r$. In terms of the asymptotic result of Eq. \eqref{eq:SrAsympt}, we expect
\begin{equation}
    \beta = \frac{f(n,q,r)}{n r^{1/2}},
\end{equation}
and indeed we find $f(n,q,r)/n r^{1/2} \simeq 0.7203$ for large values of $r$.

In the left panel of Fig.~\ref{Fig:Sn_Hamming}, we compare the exact results for the entropy obtained with Eq.~\eqref{eq:SAn_exact} for various values of $n,q,r$ (symbols) with the numerical fit of Eq.~\eqref{eq:Snfit} (solid line), and find excellent agreement. We also display the asymptotic value $f(n,q,r)/n r^{1/2} $ for $n=3$, $q=4$ and $r=600$ (dashed line).

For fixed $n,q,r$ the volume $V_\A$ of $\A$ is proportional to the area $|\partial_\A|$ of the boundary between $\A$ and $\B$, see Eq.~\eqref{eq:VADA}. Moreover, since the $d$-dependence of $|\partial_\A|$ is exponential, $|\partial_\A| \propto q^d$, the diameter $d$ scales as $\log |\partial_\A|$, up to additive and multiplicative constants. Therefore, the scaling of the entanglement entropy in Eq.~\eqref{eq:SrAsympta} corresponds to a violation of the area law by a power of the logarithm of the area, 
\begin{equation}
\label{eq:logArea12}
    S \sim |\partial_\A| \Big(\log |\partial_\A|\Big)^{-1/2}.
\end{equation}

Logarithmic violations of the area law have been observed in numerous other physical systems, such as one-dimensional quantum critical systems \cite{vidal2003entanglement,CC04} and free fermions in higher dimensions \cite{wolf2006violation,eisert2010colloquium}. However, in these cases the models are gapless and logarithmic violation tends to increase the entanglement compared to the area law, which is in stark contrast with the Hamming graph. 

The interpretation of the unusual scaling in Eq. \eqref{eq:logArea12} is that the entanglement in the Hamming graph is extremely local. The entanglement entropy per edge connecting a site in $\A$ with a site in $\B$ is of order $d^{-1/2}$ and vanishes in the large-$d$ limit. However, this effect of short-ranged entanglement is in part compensated by the fact that the number of links, or area, grows exponentially fast with~$d$, leading to an exponential amount of entanglement which is nonetheless subleading compared to the strict area law. 

\subsection{Numerics for $r\propto d$}\label{sec:numEE}

We also consider the case where $r$ grows linearly with~$d$, namely $r=(1-\delta)d$ with fixed $0< \delta < 1$. In that case, we numerically observe that the entropy scales as
\begin{equation}
\label{eq:Sdelta}
    S\Big(\bigcup_{j=1}^n \A_j\Big) = \tilde{\beta} V_\A+\dots
\end{equation}
in the limit $d\to \infty$. Here, $\tilde{\beta}$ is a constant with respect to $n,q$, but it depends on $\delta$. In the right panel of Fig.~\ref{Fig:Sn_Hamming}, we show the exact results obtained with Eq.~\eqref{eq:SAn_exact} for the entropy as a function of $d$ for $n=2$ and various values of $q,\delta$ (symbols). We do not show the results for different values of $n$, because we find that curves that only differ by the value of $n$ are almost indistinguishable. We fit the constant $\tilde{\beta}$ from Eq.~\eqref{eq:Sdelta} and find 
\begin{equation}
\label{eq:fittedTildeBeta}
  \frac{\tilde{\beta}}{(1-\delta)^{1/2}}  \simeq \left\{
\begin{array}{cc}
0.6988, & \delta=1/5, \\[.3cm]
0.7043, & \delta=2/5.
\end{array}
\right.
\end{equation}
These two numerical values are the solid lines in the right panel of Fig.~\ref{Fig:Sn_Hamming}. Let us mention that for $n=1$ and $q=2$, Eq.~\eqref{eq:Sdelta} reproduces the scaling $S \propto 2^{L}$ observed in Ref.~\cite{BCV21}.

The results for the scaling of the entanglement entropy for large $d$ but fixed $r$ and for $r=(1-\delta)d$ are mutually compatible. Indeed, comparing Eqs. \eqref{eq:Snr} and \eqref{eq:Sdelta} , we expect
\begin{equation}
\label{eq:tildeBetaPrediction}
    \frac{\tilde{\beta}}{(1-\delta)^{1/2}}=  \beta -\gamma (1-\delta) + \dots
\end{equation}
where the ellipsis indicate terms of order $(1-\delta)^2$. Injecting the numerical values of Eq.~\eqref{eq:betagamma} in Eq.~\eqref{eq:tildeBetaPrediction}, we find $ \beta -\gamma (1-\delta)\simeq 0.6981$ and $ \beta -\gamma (1-\delta)\simeq 0.7036$ for $\delta=1/5$ and $\delta=2/5$, respectively. These values are very close to the fitted ones in Eq.~\eqref{eq:fittedTildeBeta}. 

In the situation where $r \propto d$ with fixed $n,q$, we have $V_\A \propto |\partial_\A|/d $, see Eq.~\eqref{eq:VADA}. The scaling of the entanglement entropy of Eq. \eqref{eq:Sdelta} thus corresponds to a logarithmic violation of the area law of the form
\begin{equation}
    S \sim |\partial_\A| \Big(\log |\partial_\A|\Big)^{-1}.
\end{equation}
While the scaling is not identical as in the case of fixed~$r$ given in Eq. \eqref{eq:logArea12}, we also observe a logarithmic violation that tends to decrease the amount of entanglement compared to the area law. The interpretation is the same, namely the entanglement in Hamming graphs appears to get increasingly local as we approach the thermodynamic limit. A difference compared to the case of finite $r$ is that here the volume of $\A$ becomes negligible compared to the total number of sites, see Eq. \eqref{eq:aspectRatio}. It is known that the entanglement of skeletal regions with no volume is weaker than for systems with volume \cite{berthiere2022entanglement}; we observe a similar behavior here.

\section{Multipartite information}\label{sec:IV}

We use the exact result of Eq. \eqref{eq:SAn_exact} for the entanglement entropy of disjoint Hamming subgraphs to study the behavior of the mutual and the tripartite information. For simplicity, we always consider the case where $\F=\{0,1,\dots,k_0\}$ and $k_0=\lfloor d/q \rfloor$, as in Sec.~\ref{sec:r1}. However, our results readily generalize to the case of arbitrary~$\F$ if we use Eq. \eqref{eq:SLambda} instead of Eq. \eqref{eq:SAn_exact} for the entropies. 

\subsection{Exact results}

Combining Eqs. \eqref{eq:infoMut} and \eqref{eq:SAn_exact} for $n=1,2$, we find
\begin{multline}
\label{eq:I2Ex}
    I_2({\A_1:\A_2}) = \sum_{i=1}^{r} \begin{pmatrix}
    L \\ d-i-k_0
\end{pmatrix}(q-1)^{d-i-k_0}\\ \times \Big(2s(F_{i,0}^{(1)})-s(F_{i,0}^{(2)})-s(F_{i,1}^{(2)})\Big).
\end{multline}
As a consistency check, let us verify that Eq. \eqref{eq:I2Compl} holds when $\A_1$ and $A_2$ are complementary. In the context of the Hamming graph, this situation corresponds to the choices $q=2$ and $r=1$. The sum in Eq. \eqref{eq:I2Ex} simplifies to one term with $i=r=1$. With Eq. \eqref{eq:Fsimple}, we find
\begin{equation}
     I_2({\A_1:\A_2}) = 2\begin{pmatrix}
    L \\ d-1-k_0
\end{pmatrix} s(1/2),
\end{equation}
where we used $s(0)=s(1)=0$. A direct comparison with Eq. \eqref{eq:SAn_exact} yields $I_2({\A_1:\A_2})=2 S(\A_1)$, as expected. 

For the tripartite information, we combine Eqs. \eqref{eq:infoTri} and \eqref{eq:SAn_exact} for $n=1,2,3$, we find
\begin{multline}
\label{eq:I3Ex}
    I_3({\A_1:\A_2:\A_3}) = \sum_{i=1}^{r} \begin{pmatrix}
    L \\ d-i-k_0
\end{pmatrix}(q-1)^{d-i-k_0}\\ \times \Big(3s(F_{i,0}^{(1)})-3s(F_{i,0}^{(2)})-3s(F_{i,1}^{(2)})+s(F_{i,0}^{(3)})+2s(F_{i,1}^{(3)})\Big).
\end{multline}
As for the mutual information, we verify that the tripartite information in Eq.~\eqref{eq:I3Ex} is compatible with Eq. \eqref{eq:I3Compl} and vanishes when $\A_1$, $\A_2$ and $\A_3$ are complementary. This situation corresponds to $q=3$ and $r=1$. With Eq.~\eqref{eq:Fsimple}, we see that 
\begin{equation}
\begin{split}
     I_3({\A_1:\A_2:\A_3}) &= \begin{pmatrix}
    L \\ d-1-k_0 
\end{pmatrix}2^{d-1-k_0} \\ & \hspace{1.5cm}\times \Big(3s(2/3)-3s(1/3)\Big) \\
&=0
\end{split}
\end{equation}
where we used $s(x)=s(1-x)$.

\subsection{Asymptotics for finite $r$}

Using similar tools as in Sec. \ref{sec:asymptFinr}, we investigate the asymptotic behavior of the mutual information in the limit $d \to \infty$ with finite $r$. We find 
\begin{subequations}
\label{eq:I23rAsympt}
    \begin{equation}
    \label{eq:I23rAsympta}
    \begin{split}
         I_2({\A_1:\A_2}) & \sim q^{d-r} \left(\frac{1}{d} \right)^{1/2} g_2(q,r) ,\\
         I_3({\A_1:\A_2:\A_3}) &\sim q^{d-r} \left(\frac{1}{d} \right)^{1/2} g_3(q,r),
         \end{split}
    \end{equation}
with 
\begin{equation}
\label{eq:I2rAsymptb}
\begin{split}
    g_2(q,r) = & \frac{q}{\sqrt{2\pi (q-1)}} \\ & \times \sum_{i=1}^{r}   \Big(2s(F_{i,0}^{(1)})-s(F_{i,0}^{(2)})-s(F_{i,1}^{(2)})\Big)
    \end{split}
\end{equation}
and
\begin{multline}
\label{eq:I3rAsymptb}
    g_3(q,r) = \frac{q}{\sqrt{2\pi (q-1)}}  \sum_{i=1}^{r}  \Big(3s(F_{i,0}^{(1)})-3s(F_{i,0}^{(2)})\\ -3s(F_{i,1}^{(2)})+s(F_{i,0}^{(3)})+2s(F_{i,1}^{(3)})\Big)
\end{multline}
\end{subequations}
at leading order for large $d$. The match between these predictions and exact numerical calculations is extremely good, similarly as in Secs. \ref{sec:r1} and \ref{sec:asymptFinr} for the entanglement entropy, and therefore we do not illustrate it with a new figure. 

\subsection{Negligible multipartite information}

\begin{figure}
\includegraphics[width=0.45\textwidth]{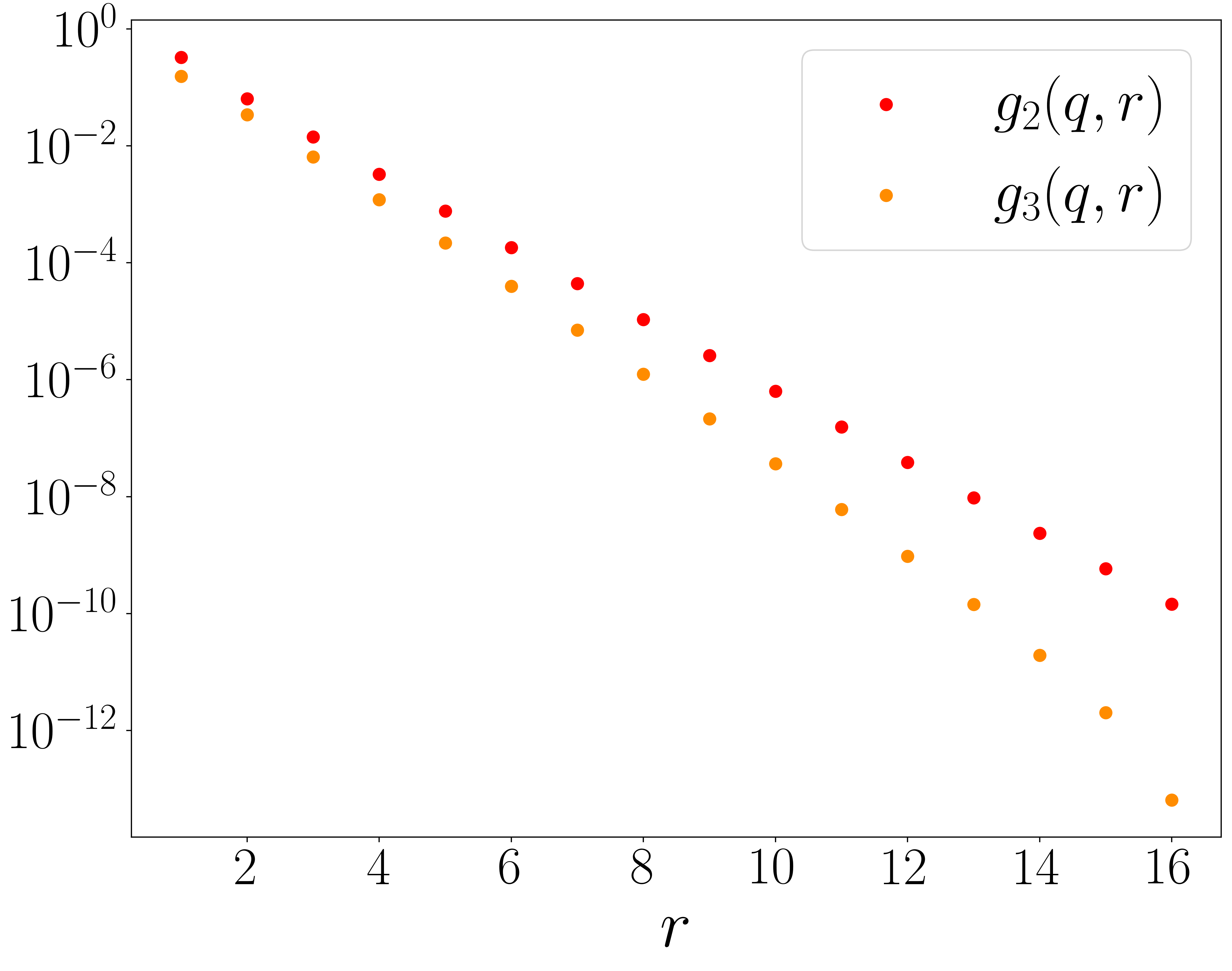}   
\caption{Functions $g_2(q,r)$ and $g_3(q,r)$ defined in Eqs. \eqref{eq:I2rAsymptb} and \eqref{eq:I3rAsymptb} for $q=5$ as a function of $r$, in logarithmic scale.} 
\label{Fig:g23}
\end{figure}

In Fig.~\ref{Fig:g23} we illustrate the behavior of $g_{2}(q,r)$ and $g_{3}(q,r)$  as a function of $r$ for fixed $q=5$. Both functions decay exponentially fast in $r$, and this behavior holds for arbitrary values of $q>2$. For $q=2$, the function $g_2(2,r)$ decays as $r^{-1/2}$ instead of exponentially, and the tripartite information is not well defined. These results indicate that for moderately large values of $r$ and~$q~>~2$, both the mutual and the tripartite information are subleading and negligible compared to the entanglement entropy. This result confirms that the entanglement in Hamming is extremely local, and hence separated regions are almost not entangled. We note that if the entanglement entropy of $n$ disjoint subgraphs were exactly proportional to $n$, we would have $S(\A_1\cup \A_2) = S(\A_1)+ S(\A_2)$, and hence $I_2({\A_1:\A_2})=I_3({\A_1:\A_2:\A_3})=0$. Here, the mutual and tripartite information do not exactly vanish precisely because of the subleading corrections that are not proportional to $n$ in the entanglement entropy, as discussed below Eq. \eqref{eq:Snr}. We stress that the near proportionality of the entanglement entropy in $n$ also holds for $n>3$, and hence we conclude that any multipartite information measure between disjoint Hamming subgraphs is negligible and subleading compared to the entanglement entropy.

Finally, we investigate the behavior of the multipartite information measures in the case where $r=(1-\delta) d$. This situation corresponds to the limit of infinite separation between the subsystems, and we thus expect mutual information measures to vanish. Indeed, the coefficients $g_2(q,(1-\delta) d)$ and $g_3(q,(1-\delta) d)$ vanish in the limit $d \to \infty$, and we find $I_2({\A_1:\A_2})=I_3({\A_1:\A_2:\A_3})=0$ for $q>2$. For $q=2$, the decay of the coefficient $g_2(2,(1-\delta) d)$ for large $d$ is slower than for $q>2$ and the mutual information does not identically vanish in the large-$d$ limit. We nonetheless have $\lim_{d \to \infty} I_2/ (q^{d \delta}d^{-1/2})=0$, namely the leading term in the expansion vanishes.

\section{Conclusion}\label{sec:V}

In this paper, we investigated multipartite entanglement and information measures for free fermions defined on Hamming graphs. We obtained the exact diagonalization of the chopped correlation matrix $C_\A$ in the case where $\A$ consists of $n$ disjoint Hamming subgraphs embedded in a larger one, see Eq. \eqref{eq:CLambda}. We stress that, while it is common to use the chopped correlation matrix to investigate entanglement in free-fermion systems, only in very rare instances can the correlation matrix be exactly diagonalized in finite size. This is thus one of the main results of this paper, and it is a non-trivial generalization of the formulas obtained in Ref. \cite{BCV21} for the case $n=1$. 

With the exact diagonalization of the chopped correlation matrix, we derived exact formulas for the entanglement entropy of disjoint blocks, as well as for the mutual and tripartite information. We focused on two models of free fermions on Hamming graphs, a nearest-neighbor hoping model and a long-range model with couplings that decay exponentially with the distance, and found that the entanglement entropy is the same in both. We investigated the asymptotic behavior of the entanglement entropy of disjoint blocks in the thermodynamic limit  $d\to \infty$ with (i)~$r=\textit{cst}$, and (ii)~$r \propto d$. In the limit (i), we found an analytic expression for the asymptotic behavior of the entropy, see Eq. \eqref{eq:SrAsympt}, and conjectured the general scaling of Eq. \eqref{eq:Snfit}. The agreement with numerical results is excellent. In the limit (ii) we mainly used numerical fits, but we argued that the conjecture of Eq. \eqref{eq:Sdelta} is compatible with the analytical results of the case (i). In both cases, the entropy scales as $S \sim  |\partial_\A| (\log |\partial_\A|)^{-a}$ with $a=1/2$ and $a=1$ for the limits (i) and (ii), respectively. These scaling correspond to logarithmic violations of the area law that tend to decrease the amount of entanglement compared to the area law. The entanglement in Hamming graphs is thus ultra local and the entanglement per boundary edge vanishes in the large-$d$ limit. Moreover, the suppression of the area law is stronger in the limit~(ii) where the subsystems have a negligible volume and become skeletal regions. Using similar methods, we also investigated the asymptotic behavior of the mutual and tripartite information and found that their leading terms are negligible compared to the entanglement entropy, and they vanish in the limit~(ii). These results for the multipartite information measures confirm that entanglement on Hamming graphs is more local than on traditional cubic lattices. This highlights the fact that the underlying geometry of a lattice model can play a non-trivial role in the physics at a thermodynamic scale.

We conclude with some directions for future research. First, it would be important to understand if other physical model display similar unusual violation of the area law as the ones we observed for Hamming graph. Second, it would be interesting to perform similar analysis on other relevant graphs, such as the Johnson graph, see Ref.~\cite{bernard2021entanglement}, and compare the behavior of multipartite entanglement.  Finally, since the Hamming graph is known to admit perfect state transfer, a natural question would be to investigate the nonequilibrium entanglement dynamics after a quantum quench and understand how perfect state transfer affects the time evolution of entanglement and multipartite information.

\begin{acknowledgements}
We thank an anonymous referee for useful comments and Pok Man Tam for interesting discussions. GP holds a CRM-ISM postdoctoral fellowship and acknowledges support from the Mathematical Physics Laboratory of the CRM. PAB holds an Alexander-Graham-Bell scholarship from the Natural Sciences and Engineering Research Council of Canada (NSERC).  
NC is supported by the international research project AAPT of the CNRS and the ANR Project AHA ANR-18-CE40-0001. The research of LV is funded in part by a Discovery Grant from NSERC.
\end{acknowledgements}

\appendix

\section{Diagonalization of $C_\A$}\label{app:CA}
In this appendix, we give the proof of Eq. \eqref{eq:CLambda}. First, we define the projectors $P_i$ as
\begin{equation}
P_i = |i\rangle \langle i |, \quad i=0,\dots,q-1,
\end{equation}
where $|i\rangle$ is a vector in $\mathbb{C}^q$ defined as in Eq. \eqref{eq:vi}. The projector on $\A$ is
\begin{equation}
\pi_\A =\big(P_0^{\otimes r} + \dots + P_{n-1}^{\otimes r}\big) \otimes (\mathbb{1}_{q \times q})^{\otimes L},
\end{equation}
where we recall that $r= d-L$. 

Second, we investigate the projector on $\F$. For simplicity, we introduce
\begin{equation}
O_i=x_i\mathbb{1}_{q \times q}+y_i J_{q \times q}
\end{equation} 
and consider the projector 
\begin{equation}
\label{eq:pid}
\pi_d(x,y) = O_1 \otimes \cdots \otimes O_{d}
\end{equation}
with $x=(x_1,\dots,x_d)$, and similarly for $y$. For a binary string $b=(b_1,b_2,\dots,b_d) \in \{0,1\}^d$ of length $d$, we define the weight function as $w(b)\equiv \sum_{i=1}^d b_i$. With these notations, the projector $\pi_\F$ is \cite{BCV21}
\begin{multline}
\label{eq:piF}
 \pi_\F = \\ \sum_{k \in \F}  \sum_{\substack{b \in \{0,1\}^d \\ w(b)=d-k}} \pi_d\left((b_1,\dots,b_d),\Big(\frac{(-1)^{b_1}}{q},\dots,\frac{(-1)^{b_d}}{q}\Big)\right).
\end{multline}

By construction, we have 
\begin{equation}
    \pi_\A |\theta_{i_1} \cdots \theta_{i_L}\rangle_{j}=|\theta_{i_1} \cdots \theta_{i_L}\rangle_{j}.
\end{equation}
Moreover, the state $|\theta_{i_1} \cdots \theta_{i_L}\rangle_{j}$ is an eigenvector of the operator $O_{r+1}\otimes \cdots \otimes O_{d}$ for any choice of $x,y$,
\begin{equation}
    O_{r+1}\otimes \cdots \otimes O_{d}|\theta_{i_1} \cdots \theta_{i_L}\rangle_{j} = \gamma_i(x,y) |\theta_{i_1} \cdots \theta_{i_L}\rangle_{j}
\end{equation}
with
\begin{equation}
    \gamma_i(x,y)= \prod_{m=1}^L (x_{r+m} +q y_{r+m} \delta_{i_m,q}).
\end{equation}

To show that $|\theta_{i_1} \cdots \theta_{i_L}\rangle_{j}$ is an eigenvector of $C_\A$, it remains to show that
\begin{multline}
\label{eq:Ppilambda}
    \big(P_0^{\otimes r} + \dots + P_{n-1}^{\otimes r}\big) \cdot \pi_{r}(x,y) |\theta_{i_1} \cdots \theta_{i_L}\rangle_{j} =\\ \lambda_j(x,y) |\theta_{i_1} \cdots \theta_{i_L}\rangle_{j}
\end{multline}
where $\pi_{r}(x,y)$ is defined as in Eq. \eqref{eq:pid} with $d \rightarrow r$. We find 
\begin{equation}
\label{eq:lambdaxy}
    \lambda_j(x,y) = \prod_{i=1}^r (x_i+y_i)+\left( \sum_{\ell=1}^{n-1}\eta_j^\ell\right)\prod_{i=1}^r y_i,
\end{equation}
where the sum over $\eta_j^\ell$ is a simple trigonometric series which simplifies to
\begin{equation}
\label{eq:trigo}
     \sum_{\ell=1}^{n-1}\eta_j^\ell =n \delta_{j,0}-1.
\end{equation}
 
 We give the proof of Eq. \eqref{eq:lambdaxy} for $r=2$ and arbitrary~$n$. For clarity, we define 
 \begin{equation}
   |\psi_n\rangle_j \equiv  \frac{1}{\sqrt{n}}\left(\sum_{\ell=0}^{n-1} \eta_j^\ell \ |\ell\rangle\otimes |\ell\rangle\right).
 \end{equation}
 The operator $\big(P_0^{\otimes 2} + \dots + P_{n-1}^{\otimes 2}\big) \cdot \pi_{2}(x,y)$ acts non-trivially on the term $|\psi_n\rangle_j$ of the vector $|\theta_{i_1} \cdots \theta_{i_L}\rangle_{j}$. The action of $\pi_{2}(x,y)$ is
 \begin{equation}
      \pi_{2}(x,y)|\psi_n\rangle_j= \frac{1}{\sqrt{n}}\sum_{\ell=0}^{n-1} \eta_j^\ell \ O_1|\ell\rangle\otimes O_2|\ell\rangle
 \end{equation}
and we have 
\begin{multline}
\label{eq:O1O2}
    O_1 |\ell\rangle\otimes O_2 |\ell\rangle = \big((x_1+y_1)(x_2+y_2)-y_1y_2\big) |\ell\rangle\otimes|\ell\rangle \\ +y_1y_2\sum_{m,m'=0}^{q-1}|m\rangle\otimes |m'\rangle.
\end{multline}
In Eq. \eqref{eq:Ppilambda}, because of the projector operators on~$\A$, the double sum over $m,m'$ in Eq. \eqref{eq:O1O2} reduces to $\sum_{m=0}^{n-1}|m\rangle\otimes |m\rangle$. We have
\begin{multline}
\label{eq:endproof}
    \big(P_0^{\otimes 2} + \dots + P_{n-1}^{\otimes 2}\big) \cdot \pi_{2}(x,y)|\psi_n\rangle_j= \\[.3cm] \big((x_1+y_1)(x_2+y_2)-y_1y_2\big)|\psi_n\rangle_j \\[.3cm] + y_1y_2 \frac{1}{\sqrt{n}}\left(\sum_{\ell=0}^{n-1} \eta_j^\ell\right)\sum_{m=0}^{n-1}|m\rangle\otimes |m\rangle.
\end{multline}
With the trigonometric identity of Eq. \eqref{eq:trigo}, we conclude that Eq. \eqref{eq:endproof} is exactly Eq. \eqref{eq:Ppilambda} with the eigenvalue of Eq. \eqref{eq:lambdaxy} for $r=2$. The generalization for arbitrary $r$ is direct.

Combining the results from the previous paragraphs, we thus find
\begin{multline}
    \pi_\A \pi_d(x,y)\pi_\A  |\theta_{i_1} \cdots \theta_{i_L}\rangle_{j} = \\ \lambda_j(x,y) \gamma_i(x,y) |\theta_{i_1} \cdots \theta_{i_L}\rangle_{j}.
\end{multline}
Using this result and Eq. \eqref{eq:piF} to express $\pi_\F$ in terms of $\pi_d(x,y)$, we conclude that $ |\theta_{i_1} \cdots \theta_{i_L}\rangle_{j}$ is an eigenvector of the chopped correlation matrix, with the eigenvalue given in Eq. \eqref{eq:CLambda}.

\providecommand{\href}[2]{#2}\begingroup\raggedright\endgroup

\end{document}